\documentclass[conference]{IEEEtran}
\IEEEoverridecommandlockouts
\usepackage{cite}
\usepackage{amsmath,amssymb,amsfonts}
\usepackage{algorithmic}
\usepackage{graphicx}
\usepackage{textcomp}
\usepackage{xcolor}
\usepackage[font=small,labelfont=bf]{caption}
\def\BibTeX{{\rm B\kern-.05em{\sc i\kern-.025em b}\kern-.08em
    T\kern-.1667em\lower.7ex\hbox{E}\kern-.125emX}}
\begin{document}

\title{Frequency-based Automated Modulation Classification in the Presence of Adversaries  \\
\thanks{This project was supported in part by the Naval Surface Warfare Center Crane Division and in part by the National Science Foundation (NSF) under grants CNS1642982  and CCF1816013.}
}


\author{ }
\author{Rajeev Sahay, Christopher G. Brinton, and David J. Love \\ School of Electrical and Computer Engineering, Purdue University \\ \{sahayr,cgb,djlove\}@purdue.edu}

\maketitle

\begin{abstract}

Automatic modulation classification (AMC) aims to improve the efficiency of crowded radio spectrums by automatically predicting the modulation constellation of wireless RF signals. Recent work has demonstrated the ability of deep learning to achieve robust AMC performance using raw in-phase and quadrature (IQ) time samples. Yet, deep learning models are highly susceptible to adversarial interference, which cause intelligent prediction models to misclassify received samples with high confidence. Furthermore, adversarial interference is often \emph{transferable}, allowing an adversary to attack multiple deep learning models with a single perturbation crafted for a particular classification network. In this work, we present a novel receiver architecture consisting of deep learning models capable of withstanding transferable adversarial interference. Specifically, we show that adversarial attacks crafted to fool models trained on time-domain features are not easily transferable to models trained using frequency-domain features. In this capacity, we demonstrate classification performance improvements greater than 30\% on recurrent neural networks (RNNs) and greater than 50\% on convolutional neural networks (CNNs). We further demonstrate our frequency feature-based classification models to achieve accuracies greater than 99\% in the absence of attacks. 

\end{abstract}

\begin{IEEEkeywords}
Adversarial attacks, automatic modulation classification, machine learning, privacy, security
\end{IEEEkeywords}

\section{Introduction}

\begin{figure*}[htb] 
	\centering
	\includegraphics[width=2.0\columnwidth]{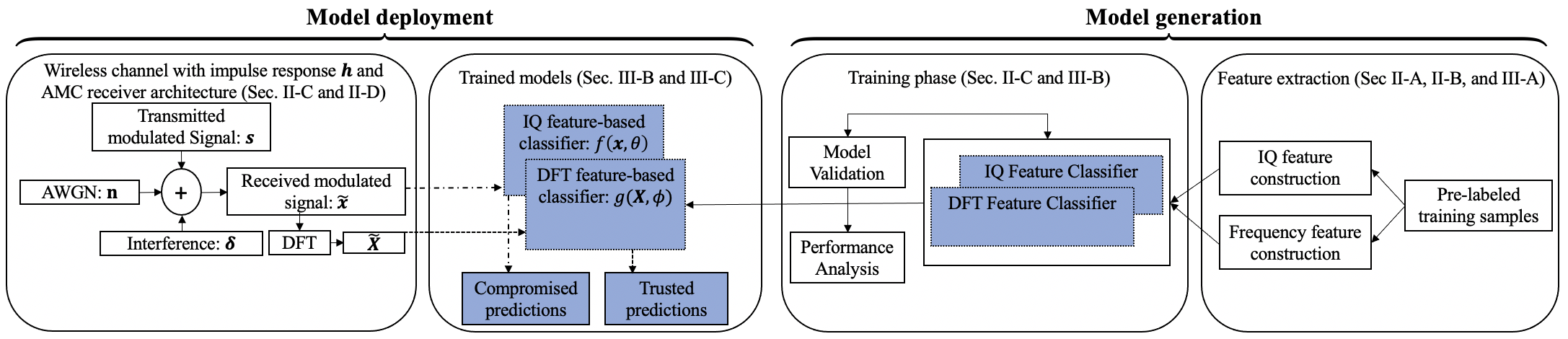}
	\caption{Our AMC system model with adversarial interference. The shaded blocks correspond to our time and frequency-domain classifiers.}
	\label{block_diagram}
\end{figure*}

\IEEEPARstart{T}{he} recent exponential growth of wireless traffic has resulted in a crowded radio spectrum, which, among other factors, has contributed to reduced mobile efficiency. With the number of devices requiring wireless resources projected to continue increasing, this inefficiency is expected to present large-scale challenges in wireless communications. Automatic modulation classification (AMC), which is a part of cognitive radio technologies, aims to alleviate the inefficiency induced in shared spectrum environments by dynamically extracting meaningful information from massive streams of wireless data. Traditional AMC methods are based on maximum-likelihood (ML) approaches \cite{ml_review}, which consist of deriving statistical decision boundaries using hand-crafted features to discern various modulation constellations. More recently, deep learning (DL) has become a popular alternative to ML methods for AMC, since it does not require manual feature engineering to attain robust classification performance \cite{amc_dl1}. 


Despite their robust AMC performance, however, deep learning models are highly susceptible to adversarial attacks \cite{intriguing_props}, which introduce additive wireless interference into transmitted RF signals to induce high-confidence misclassifications on well-trained deep learning models. In addition to degrading the classification performance of a particular targeted model, adversarial attacks are also transferable to other classification networks that are trained to perform the same task as the targeted classifier \cite{transfer1}. As a result, an adversary can degrade the performance of several deep learning models simultaneously, thus reducing spectrum efficiency and compromising secure communication channels.



In this work, we develop a novel AMC method that is capable of mitigating the effects of transferable adversarial attacks. Specifically, our method learns on frequency domain-based features, as opposed to in-phase and quadrature (IQ) time-domain features, which are traditionally used for deep learning-based AMC. After quantifying the model's performance in the absence of adversarial interference, we consider a wireless channel compromised by an adversary aiming to induce an erroneous modulation constellation prediction at the receiver by injecting interference into the transmitted signal. Although the interference degrades the classification performance of the model trained on IQ features, the frequency feature-based model significantly increases the probability of correctly classifying the perturbed signal, thus mitigating the effects of transferable adversarial interference. 

\textbf{Related Work:} The susceptibility of deep learning-based AMC models to adversarial attacks has been demonstrated in prior work \cite{adv_filters,amc_adv_atk1,amc_adv_atk2}. Such attacks have been found to be more efficient than traditional jamming attacks applied in communication networks \cite{jamming} and, as a result, present challenges for deep learning deployment in autonomous wireless channels \cite{hurdle1}. Yet, limited work has been conducted in exploring the degree to which AMC models are susceptible to interference. Few defenses have been proposed to mitigate the effects of wireless adversarial interference \cite{autoencoder_defense}, and to the best of our knowledge, no work has explored the extent to which adversarial attacks are transferable between domains (although various domains for classification have been investigated \cite{fft_features}). On the other hand, several defenses have been proposed for defending deep learning image classifiers from adversarial attacks, with no method generally accepted as a robust solution \cite{review1}. Nonetheless, even considering the adoption of image classification defenses for AMC is difficult due to the differing constraints placed on the adversary in both settings (e.g., transmit power budget, SNR degradation, visual perceptibly, etc.). In this work, we address this challenge by proposing a novel AMC methodology, which allows us to quantify the extent of adversarial transferability in a wireless channel with real-world communication constraints. 



\textbf{Summary of Contributions:} The main contributions of this work are as follows: 

\begin{enumerate}
    \item \textbf{A novel signal receiver architecture for AMC} (Sec. II-B, II-C, and III-B): We model and develop a robust AMC module consisting of both frequency-based and IQ-based deep learning architectures. 
    
    \item \textbf{Resilience to time domain adversaries} (Sec. II-D and III-C): We demonstrate that, although an adversary may be able to degrade the classification performance on a time-domain model, their attacks are not well-transferable to our models trained using frequency features. 
    
    \item \textbf{Best architecture to offset adversary} (Sec. III-C): Our results show that, out of several deep learning architechures, convolutional neural networks (CNNs) have the fastest training times and mitigate the classifier degradation to the greatest extent. 
    
\end{enumerate}

\section{Our AMC Methodology}

In this section, we outline the wireless channel we consider for AMC as well as the assumptions about the knowledge level of the transmitter, receiver, and adversary. We describe two ways we represent the received signal (Sec. II-A and II-B) followed by the machine learning models we employ for AMC (Sec. II-C). Finally, we describe perturbation methods performed by the adversary to induce misclassifications on the trained models (Sec. II-D). Our overall AMC system model is shown in Fig. \ref{block_diagram}. 
\subsection{Signal Modeling}

We consider a wireless channel consisting of a transmitter, which is aiming to send a modulated signal, and a receiver, whose objective is to perform AMC on the obtained waveform and realize its modulation constellation. Specifically, at the transmitter, we consider an underlying data source, $\mathbf{s} = [s[0],\ldots,s[\ell - 1]]$, which is modulated using one of $C$ modulation constellations chosen from a set, $\mathcal{S}$, of possible modulation schemes, with each scheme having equal probability of selection. $s[k]$ is the (scalar) value wirelessly transmitted at time $k$. At the receiver, the collected waveform at each time instance is modeled by
\begin{equation} \label{x(t)}
    x[k] = \sqrt{\rho}(\mathbf{s} * \mathbf{h})[k] + n[k],
\end{equation} 
\noindent $n[k]$ represents complex additive white Gaussian noise (AWGN) at time $k$ distributed as $\mathcal{CN}(0,1)$, $\sqrt{\rho}$ denotes the SNR (known at the receiver), $*$ denotes convolution, and $\mathbf{h}$ captures the wireless channel's impulse response. $\mathbf{h}$ also includes radio imperfections such as sample rate offset (SRO), center frequency offset (CFO), and selective fading, none of which are known to the receiver. Furthermore, we assume that the receiver has no knowledge about the channel model or the distribution of $\mathbf{h}$. This general setting motivates an AMC solution using a data-driven approach, as presented in this work, in which the true modulation constellation of the received signal is estimated from a model trained on a collection of pre-existing labeled signals. 


\subsection{Domain Transform}

At the receiver, we model $\mathbf{x} = [x[0],\ldots,x[\ell - 1]]$ using its frequency components obtained from the discrete Fourier transform (DFT). Specifically, the $p^{\text{th}}$ component of the DFT of $\mathbf{x}$ is given by 
\begin{equation} \label{dft}
    X[p] = \sum_{k=0}^{\ell - 1} x[k] e^{-\frac{j2\pi}{\ell}pk}, 
\end{equation}
\noindent where $\mathbf{X} = [X[0],\ldots,X[\ell - 1]]^T$ contains all frequency components of $\mathbf{x}$. We are interested in comparing the efficacy of AMC learning based on $\mathbf{x}$ and $\mathbf{X}$ as feature representations of the input signal. 
Although both signal representations are complex (i.e., $\mathbf{x}, \mathbf{X} \in \mathbb{C}^{\ell}$), we represent all signals as two-dimensional reals, using the real and imaginary components for the first and second dimension, respectively, in order to utilize all signal components during classification. Thus, we represent all time and frequency domain features as real-valued matrices $\mathbf{x}, \mathbf{X} \in \mathbb{R}^{\ell \times 2}$. 
\subsection{Deep Learning Architectures}

In this work, we consider the effectiveness of different deep learning architectures for AMC under IQ and frequency features as model inputs. In general, we denote a trained deep learning classifier, parameterized by $\theta$, as $f(\cdot, \theta): \mathbb{R}^{\ell \times 2} \rightarrow \mathbb{R}^{C}$. This calculates the likelihood, $\hat{\mathbf{y}}$, of an input signal consisting of IQ features, $\mathbf{x}$, belonging to each of the $C$ modulation constellations. From $\hat{\mathbf{y}}$, the predicted modulation constellation is given by ${\arg\max}_{i = 1, \ldots, C} \hspace{0.5mm} \hat{y}_{i}$. Similarly, we denote a deep learning classifier trained using the DFT of the input signal, $\mathbf{X}$, parameterized by $\phi$, as $g(\cdot, \phi): \mathbb{R}^{\ell \times 2} \rightarrow \mathbb{R}^{C}$, which is trained to perform the same classification task as $f(\cdot, \theta)$ but using the frequency features of $\mathbf{x}$ to comprise the input signal. We analyze the classification performance using the aforementioned signal representations on four common AMC deep learning architectures: the fully connected neural network (FCNN), the convolutional neural network (CNN), the recurrent neural network (RNN) and the convolutional recurrent neural network (CRNN). Each architecture consists of a set of layers and a set of neurons per layer (referred to as units). The specific differences of layer interactions in each considered model are described below. 
For each model, we apply the ReLU non linearity in its hidden layers, given by $\sigma(a) = \max\{0, a\}$, and a $C$-unit softmax output layer given by
\begin{equation}
    \sigma(\mathbf{a})_{i} = \frac{e^{a_{i}}}{\sum_{j=1}^{C} e^{a_{j}}},
\end{equation}
where $i = 1,\ldots,C$ for input vector $\mathbf{a}$. This normalization allows a probabilistic interpretation of the model's output predictions. 

\textbf{FCNN}: Our FCNN consists of three hidden layers with 256, 128, and 128 units, respectively. 
The output of a single unit, $u$, is given by
\begin{equation}
    {\sigma}\big{(}\sum_{i} w_{i}^{(u)}\cdot a_i + b\big{)},
\end{equation}
\noindent where $\sigma(\cdot)$ is the activation function, $\mathbf{w} = [w_1, \ldots, w_n]$ is the weight vector for unit $u$ estimated from the training data, $\mathbf{a} = [a_1, \ldots, a_n]$ is the vector containing the outputs from the previous layer (or the model inputs in the first layer), and $b$ is a threshold bias. Each hidden layer applies a 20\% dropout rate during training.

\textbf{CNN}: The CNN is comprised of two convolutional layers consisting of 256 and 64 feature maps (each with 20\% dropout), respectively, followed by a 128-unit fully connected layer. The output of each feature map in the convolutional layer is given by  
\begin{equation}
    {\sigma}\big{(} \mathbf{v} * \mathbf{a} + b\big{)},
\end{equation}
where $\mathbf{v}$ is the filter kernel whose parameters are estimated during training, and $\mathbf{a}$ is the output from the preceding layer. Our model uses a $2 \times 5$ and $1 \times 3$ kernel for the first and second convolutional layers, respectively. 

\textbf{RNN}: The RNN is comprised of a 75-unit long-short-term-memory (LSTM) \cite{lstm} layer followed by a 128-unit ReLU fully connected layer. Each LSTM unit implements three gates for learning. \emph{Input gates} prevent irrelevant features from entering the recurrent layer while \emph{forget gates} eliminate irrelevant features altogether. \emph{Output gates} produce the LSTM layer output, which is inputted into the subsequent network layer. The gates are used to recursively calculate the internal state of the cell, denoted by $\mathbf{z}_{c}^{(t)}$ at time $t$ for cell $c$, at a specific recursive iteration, called a time instance, which is then used to calculate the cell output given by
\begin{equation}
    \mathbf{q}^{(t)} = \text{tanh}(\mathbf{z}_{c}^{(t)})\sigma(\mathbf{p}^{(t)}), 
\end{equation}
where $\mathbf{p}^{(t)}$ is the parameter obtained from the output gate and $\sigma(\cdot)$ is the logistic sigmoid function given by $\sigma(p_{i}^{t}) = 1 / (1 + e^{-p_{i}^{t}})$ for the $i^{\text{th}}$ element in $\mathbf{p^{(t)}}$. 

\textbf{CRNN}: Lastly, we consider a CRNN comprised of two convolutional layers (containing 128 and 64 feature maps with $2 \times 5$ and $1 \times 3$ kernels, respectively) followed by a 32-unit LSTM layer.

Unless otherwise noted, each model is trained using the Adam optimizer \cite{adam}, 75 epochs, a batch size of 64, and the categorical cross entropy loss function given at the output by 
\begin{equation} \label{single_cost}
    \mathcal{L}_{n} = \sum_{j=1}^{C} y_{j} \text{log}(\hat{y}_{j})
\end{equation}
for each sample $n$ and 
\begin{equation}
    \mathcal{L} = -\frac{1}{N}\sum_{n=1}^{N} \mathcal{L}_{i},
\end{equation}
\noindent over the entire training set $n = 1,\ldots, N$, where $y_j = 1$ if the ground truth label of the sample is modulation class $j$ and $y_j = 0$ otherwise. 

\subsection{Adversarial Interference}

In addition to the transmitter and receiver, our considered communication network also consists of an adversary, whose objective is to induce a misclassification on the trained AMC model. 
The adversary will perturb the received signal by injecting wireless interference, which we will denote $\pmb{\delta}: \pmb{\delta} \in \mathbb{R}^{\ell \times 2}$, into $\mathbf{x}$ during transmission. For a given design of $\pmb{\delta}$, the resulting signal that arrives at the receiver will be
\begin{equation}
    \tilde{\mathbf{x}} = \mathbf{x} + \pmb{\delta},
\end{equation}
where $\tilde{\mathbf{x}} = \mathbf{x}$ in the absence of an attack (i.e., when $\pmb{\delta} = 0$). We consider a limited knowledge level threat model where the adversary knows the architecture and parameters of $f(\cdot, \theta)$ but is blind to $g(\cdot, \phi)$. This constraint mimics a real-world wireless channel where an adversary may not have complete knowledge of the underlying system under attack and thus restricts the adversary to injecting an attack in the time-domain, where traditional AMC features are constructed from. 

The adversary's objective is to inject $\pmb{\delta}$ to change the classification of $\mathbf{x}$ using the least amount of power possible (to evade detection caused by higher powered adversarial interference \cite{adv_det}), thus constraining the power of the perturbation to 
\begin{equation}
    ||\pmb{\delta}||_{2}^{2} \leq P_{T}, 
\end{equation}
where $P_{T}$ is the total power budget available to the adversary for instantiating an attack. In this work, we study two particular methods to inject adversarial interference: the fast gradient sign method (FGSM) \cite{fgsm}, in which the adversary exhausts its total power budget on a single step attack, and the basic iterative method (BIM) \cite{bim}, in which the adversary iteratively uses a fraction of its attack budget resulting in a more powerful attack at the cost of higher computational overhead. 

\textbf{FGSM}: In this case, the adversary adds an $l_{2}$-bounded perturbation to the transmitted signal in a single step exhausting the power budget. Formally, the $n^{\text{th}}$ perturbed received signal is given by 
\begin{equation} \label{l2_fgsm}
    \tilde{\mathbf{x}}_{n} = \mathbf{x}_{n} + P_{T} \frac {\nabla_{\mathbf{x}} \mathcal{L}_{n}(\mathbf{x}_{n}, \mathbf{y}_{n}, \theta)} {||\nabla_{\mathbf{x}} \mathcal{L}_{n}(\mathbf{x}_{n}, \mathbf{y}_{n}, \theta)||_{2}},
\end{equation}
\noindent where $\mathcal{L}$ refers to the cost function of $f(\cdot, \theta)$ in (\ref{single_cost}). Adding a perturbation in the direction of the cost function's gradient behaves as performing a step of gradient ascent, thus increasing the classification error on the perturbed sample. We explore the effects of various bounds on $P_{T}$ in Section III-C. 

\textbf{BIM} The BIM is an iterative extension of the FGSM. Specifically, in each iteration, a smaller $l_{2}$-bounded perturbation, $\alpha < P_{T}$, is added to the transmission, and the optimal direction of attack (the direction of the gradient) is recalculated. Formally, the perturbation on iteration $k + 1$ for the $n^{\text{th}}$ sample is calculated as
\begin{equation}
    \tilde{\mathbf{x}}_{n}^{(k+1)} = \mathbf{x}_{n}^{(k)} + \text{clip}_{P_{T}}\bigg{(}\alpha \frac {\nabla_{\mathbf{x}} \mathcal{L}_{n}(\mathbf{x}_{n}^{(k)}, \mathbf{y}_{n}, \theta)} {||\nabla_{\mathbf{x}} \mathcal{L}_{n}(\mathbf{x}_{n}^{(k)}, \mathbf{y}_{n}, \theta)||_{2}}\bigg{)},
\end{equation}
\noindent where the \texttt{clip} function is defined to ensure that the additive perturbation based on $\alpha$ in each iteration remains within the adversary's power budget. 


\section{Results and Discussion}

In this section, we conduct an empirical evaluation of our method. First, we overview the dataset that we use (Sec. III-A). Next, we present the efficacy of using frequency features for classification in the absence of any adversarial interference (Sec. III-B). Finally, we demonstrate the resilience of our trained models to transferable adversarial attacks instantiated in the time domain (Sec. III-C). 
\vspace{-0.2cm}
\subsection{Dataset and Evaluation Setup}

We employ the GNU RadioML2016.10B dataset \cite{dataset} for our analysis. Each signal in the dataset, $\mathbf{x}_{n}$, has an SNR of 18 dB, is normalized to unit energy, and consists of a 128-length observation window modulated according to a certain digital constellation, $\mathbf{y}_n$. We focus on the following four modulation schemes: CPFSK, GFSK, PAM4, and QPSK. Each constellation set contains 6000 examples for a total of 24000 signals. In each experiment, we employ a 70/15/15 training/validation/testing dataset split, where the training and validation data are used to estimate the parameters of $f(\cdot, \theta)$ and $g(\cdot, \phi)$, and the testing dataset is used to evaluate each trained model's susceptibility to adversarial interference and transferability to resilience. In particular, the validation set is used to tune the model parameters using unseen data during the training process whereas the testing set is used to measure the performance of the fine-tuned model. We denote the training, validation, and testing datasets, consisting of either time-domain IQ points or frequency-domain feature components, as $\mathcal{X}_{tr}^{t}$, $\mathcal{X}_{va}^{t}$, $\mathcal{X}_{te}^{t}$, $\mathcal{X}_{tr}^{\omega}$, $\mathcal{X}_{va}^{\omega}$, and $\mathcal{X}_{te}^{\omega}$, respectively.  

\subsection{Model Convergence Rate and Accuracy}

\begin{figure}[h] 
	\centering
	\includegraphics[width=1.0\columnwidth]{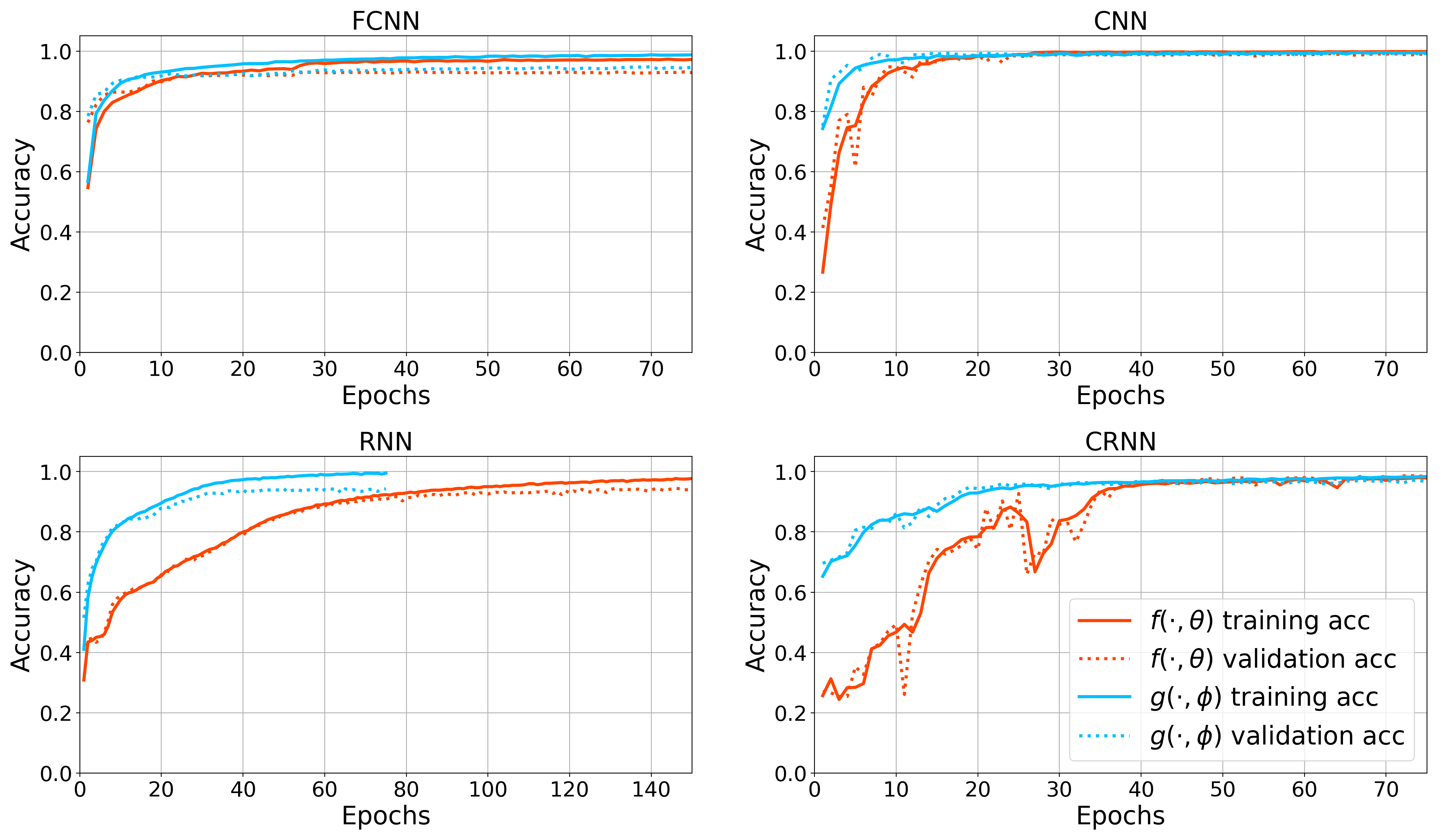}
	\caption{The model training performance of each considered AMC architecture on the corresponding training and validation sets. We see that the frequency-based features $g(\cdot, \phi)$ outperform the time domain features $f(\cdot, \theta)$ in terms of training convergence and validation accuracy for each deep learning architecture. The CNN results in the fastest convergence and highest accuracy for both $f(\cdot, \theta)$ and $g(\cdot, \phi)$.}
	\label{iq_frq_perf}
\end{figure}



\begin{table} 
\small
\caption{The testing accuracy of each considered model on $\mathcal{X}_{te}^{(\cdot)}$. The CNN outperforms every other considered model (although the CRNN delivers equivalent accuracy, it is achieved with a longer training time in Fig. \ref{iq_frq_perf} compared to the CNN). \label{model_acc}}
\centering
\begin{tabular}{c c c} 
\centering 
Model & Input Features & Accuracy \\
\hline
FCNN & IQ &  92.25\%\\
FCNN & Frequency & 92.42\% \\
CNN & IQ & 98.92\% \\
CNN & Frequency &  99.19\% \\
RNN & IQ &  93.78\%\\
RNN & Frequency & 92.67\% \\
CRNN & IQ & 98.28\% \\
CRNN & Frequency &  99.03\% \\

\hline

\end{tabular}
\end{table}

We begin by evaluating the performance of both $f(\cdot, \theta)$ and $g(\cdot, \phi)$ in the absence of adversarial interference. In Fig. \ref{iq_frq_perf}, we plot the evolution of the classification accuracy across training epochs achieved by each deep learning architecture on the training and validation sets. In contrast to using IQ training features, we see that each model trained using our proposed frequency feature-based input outperforms its time domain counter-part model. For example, the RNN trained on frequency components achieves an accuracy of 93.4\% on its corresponding validation dataset in 75 training epochs whereas the same architecture trained on $\mathcal{X}_{tr}^{t}$ requires 150 epochs to converge to a validation accuracy of 93.9\%. Furthermore, the CRNN also converges in fewer epochs when using frequency-based features in comparison to IQ features. We also see in Fig. 2 that the CNN obtains the best performance overall. IQ features present more challenges during training on the FCNN, RNN, and CRNN compared to the CNN. Specifically, the FCNN results in slight overfitting to the training data, the RNN fails to converge on a validation accuracy greater than 94\%, and the CRNN presents instability during optimization requiring a longer number of training epochs before convergence. The CNN, on the other hand, entails almost no degree of overfitting while converging in substantially less epochs compared with the RNN and CRNN models.

Each trained model's accuracy on its corresponding testing set is shown in Table \ref{model_acc}. Among all eight considered models, the CNN trained on frequency features, as proposed in this work, achieves the highest testing accuracy, as well as the fastest convergence rate in Fig. \ref{iq_frq_perf}. Specifically, this model results in nearly no overfitting between $\mathcal{X}_{tr}^{\omega}$ and $\mathcal{X}_{va}^{\omega}$, unlike either FCNN, while converging in nearly 10 epochs unlike the CNN trained on IQ features. Although the CRNN, in both cases, results in robust classification performance, the higher number of epochs required by these models results in substantially higher computational overhead (e.g., the the CNN achieves a three-fold improvement per epoch over the CRNN). \emph{Therefore, our proposed CNN trained using frequency features is the most desirable model in terms of classification performance, training time, and computational efficiency.}


\subsection{Model Resilience to Adversarial Interference}

\begin{figure}[t] 
	\centering
	\includegraphics[width=1.0\columnwidth]{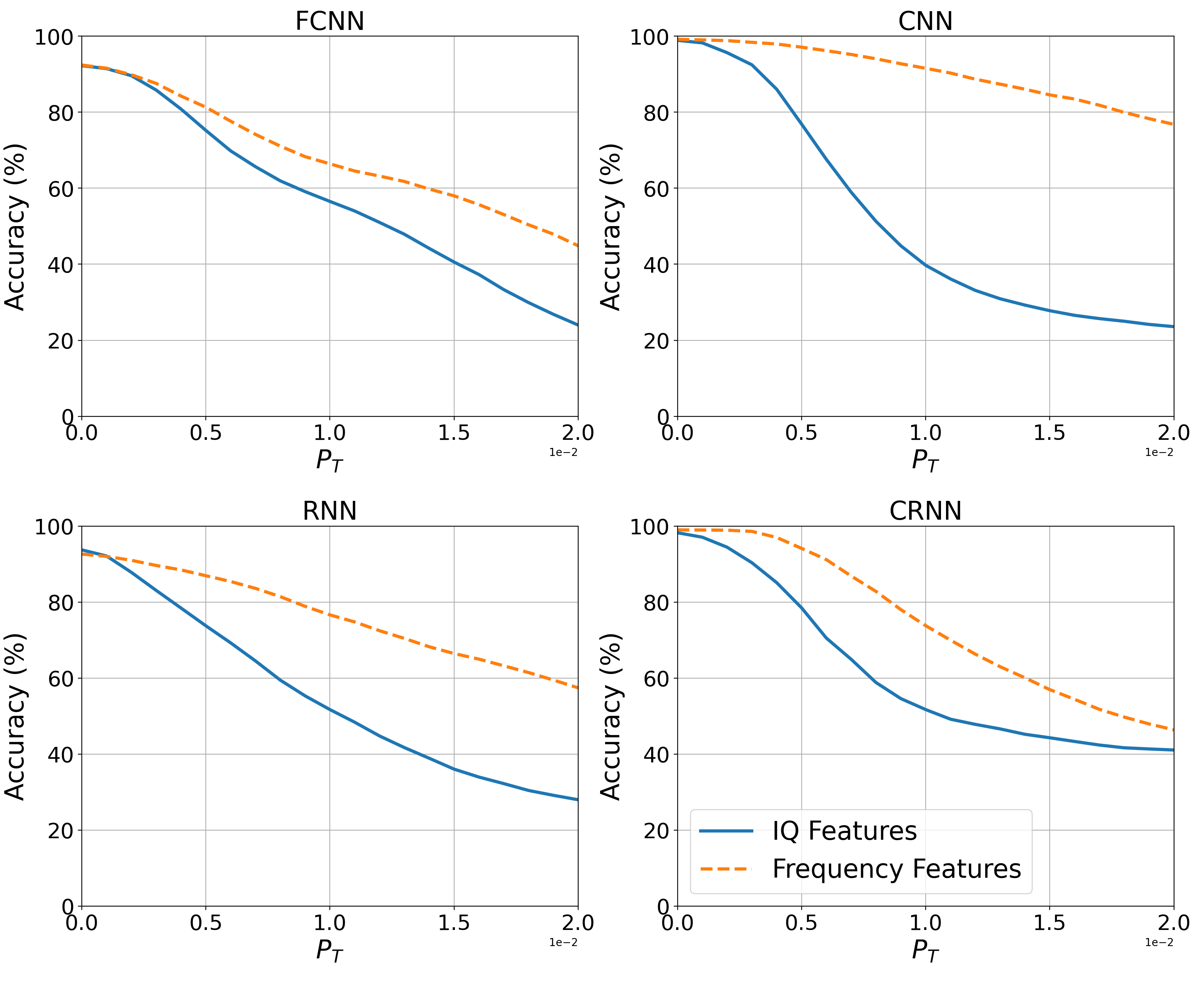}
	\caption{The transferability of the FGSM attack from $f(\cdot, \theta)$ to $g(\cdot, \phi)$. The CNN mitigates the effects of the attack to the greatest extent with a performance improvement of 53.23\% when the adversary exhausts the total perturbation budget.}
	\label{fgsm_plots}
\end{figure}

\begin{figure}[t] 
	\centering
	\includegraphics[width=1.0\columnwidth]{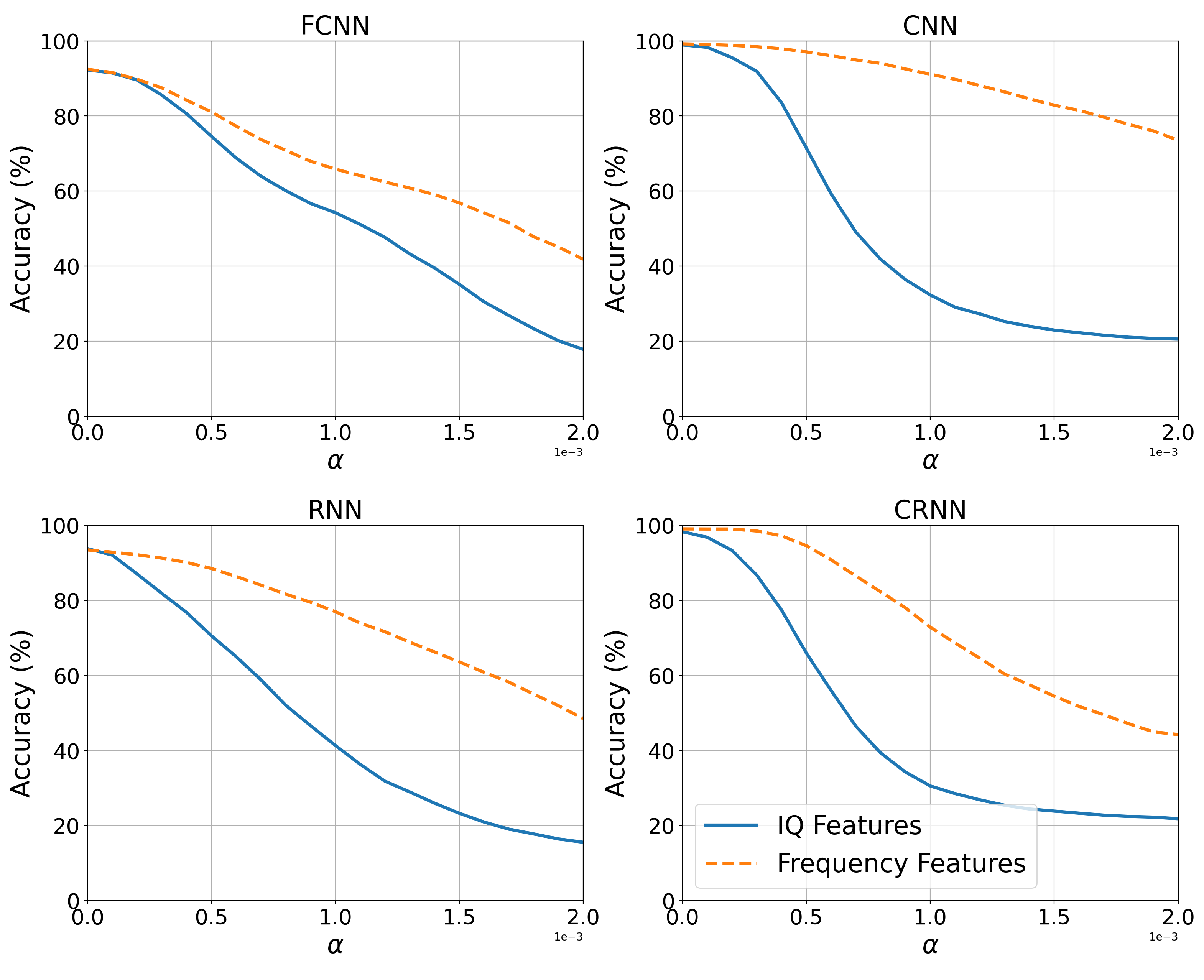}
	\caption{The transferability of the BIM attack from $f(\cdot, \theta)$ to $g(\cdot, \phi)$. Similar to the FGSM attack, the CNN displays the strongest resilience to transferability with a performance improvement of up to 52.91\%.}
	\label{bim_plots}
\end{figure}

We now evaluate the ability of an adversarial attack instantiated in the time domain to affect our frequency domain-based AMC methodology. We begin by considering the FGSM attack where we restrict $P_{T} \leq 0.0200$ (corresponding to $2\%$ additive power, which effectively degrades time domain model performance). Fig. \ref{fgsm_plots} depicts the robustness of each considered model for various levels of injected interference. We see that $g(\cdot; \phi)$ improves the classification accuracy of each model in the presence of an attack on time domain feature-based classifiers. In particular, the average accuracy improvement for the FCNN, CNN, RNN, and CRNN is 10.77\%, 38.32\%, 20.61\%, and 13.26\%, respectively, across the range of $P_{T}$. The ability of the CNN and RNN to withstand attacks to the highest degree indicates their increased resilience to transferable adversarial interference. 

The effect of the BIM adversarial attack is consistent with the response of the FGSM attacks. For the BIM attacks, we used ten iterations of different $\alpha$-bounds with $P_{T} = 0.0200$. 
As shown in Fig. \ref{bim_plots}, the attack instantiated on time domain features is significantly mitigated on each considered model when the frequency domain is used for classification. The FCNN, CNN, RNN, and CRNN experience average improvements of 12.99\%, 42.16\%, 27.31\%, and 27.33\%, respectively, for $\alpha \in [0.000, 0.002]$. \emph{Thus, as shown by the instantiation of both considered attacks, the transferability of adversarial interference is mitigated to the greatest extent when using the CNN as the underlying classification model.}

\begin{figure}[t] 
	\centering
	\includegraphics[width=1.0\columnwidth]{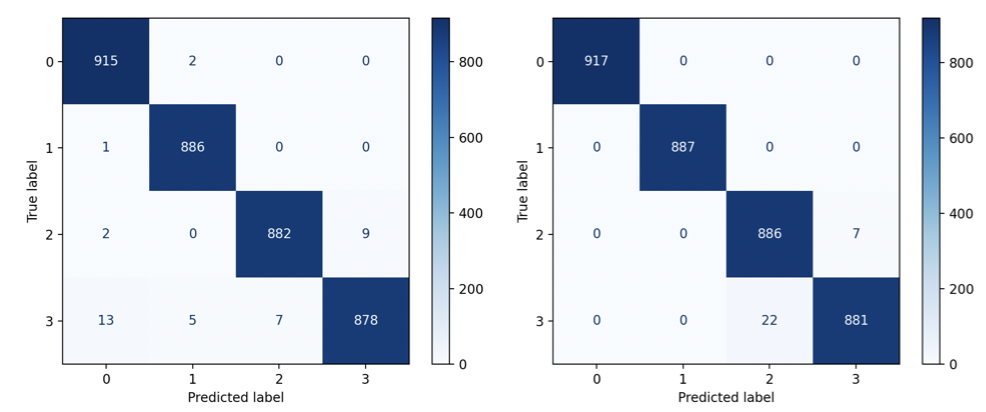}
	\caption{The confusion matrices of the CNN's predictions with no interference, using IQ features (left) and frequency features (right). The performance for both feature representations is equivalent.}
	\label{baseline_conf_matx}
\end{figure}

\begin{figure}[t] 
	\centering
	\includegraphics[width=1.0\columnwidth]{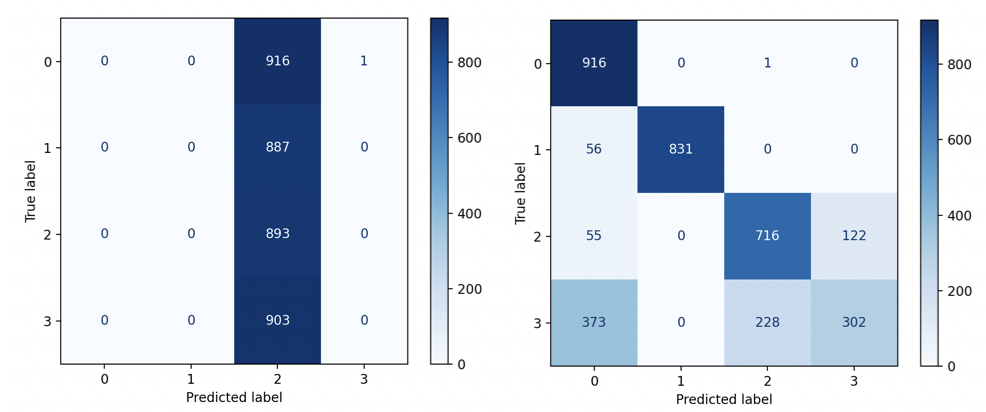}
	\caption{The confusion matrices of the CNN classifier with the FGSM perturbation, using IQ features (left) and frequency features (right). The frequency feature-based model is able to significantly mitigate the effects of the interference induced on the IQ features.}
	\label{fgsm_conf_matx}
\end{figure}

\begin{figure}[t] 
	\centering
	\includegraphics[width=1.0\columnwidth]{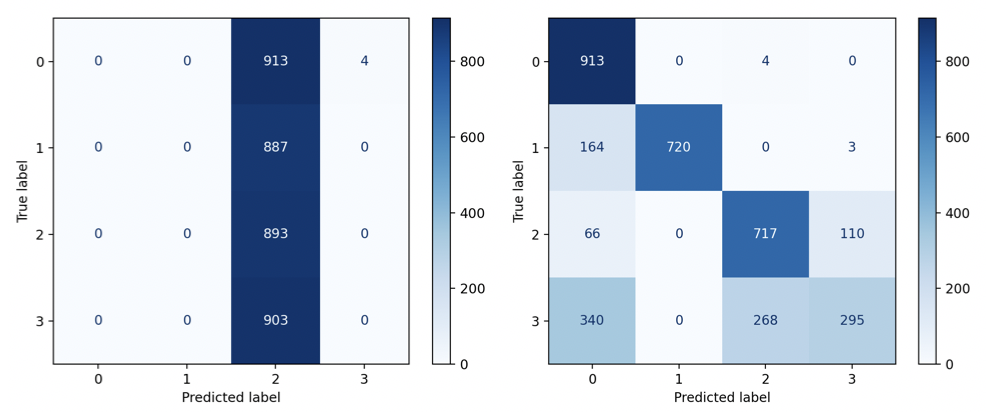}
	\caption{The confusion matrices of the CNN's predictions with the BIM attack, using IQ features (left) and frequency features (right). Similar to the FGSM attack, the CNN trained using frequency features significantly mitigates the effects of time domain feature-based perturbations.}
	\label{bim_conf_matx}
\end{figure}

We analyze the performance of the CNN model more closely, both in the presence and absence of interference, in Figs. \ref{baseline_conf_matx}-\ref{bim_conf_matx}. The labels \{0, 1, 2, 3\} correspond to the constellations \{CPFSK, GFSK, PAM4, QPSK\}. As shown in Fig. \ref{baseline_conf_matx}, both time and frequency features deliver robust AMC performance in the absence of adversarial interference with classification rates of 98.92\% and 99.19\%, respectively. However, the classification rate in the time domain drops to a mere 23.58\% and 20.56\% when the FGSM and BIM perturbations are employed, respectively (where the total perturbation budget is exhausted in both cases). As shown in Figs. \ref{fgsm_conf_matx} and \ref{bim_conf_matx}, the adversarial interference pushes the majority of signals within the classification decision boundaries of the PAM4 constellation. This is largely due to the nature of the untargeted attack in which the adversary's sole objective is to induce misclassification without targeting a specific misclassified prediction. The CNNs trained on frequency features, however, show significant improvements in classifying both FGSM and BIM perturbed signals with accuracies of 76.81\% and 73.47\% corresponding to classification accuracy improvements of 53.23\% and 52.91\%, respectively. The frequency domain-based models correctly classify a majority of CPFSK and GFSK modulation schemes, with the largest incongruency being between PAM4 and QPSK. 

\section{Conclusion and Future Work}

Deep learning has recently been proposed as a robust method to perform automatic modulation classification (AMC). Yet, deep learning AMC models are vulnerable to adversarial interference, which can alter a trained model's predicted modulation constellation with very little input power. Furthermore, such attacks are transferable, which allows the interference to degrade the performance of several classifiers simultaneously. In this work, we developed a novel wireless transmission receiver architecture, consisting of a frequency domain feature-based classification model, which is capable of mitigating the transferability of adversarial interference. Specifically, we showed that our proposed frequency-feature based deep learning classifiers are resilient to transferable adversarial interference instantiated on traditional time-domain in-phase and quadrature (IQ) feature-based models. The convolutional neural network (CNN), in particular, demonstrated the most robust classification performance in the absence of an attack, along with the highest resilience to withstand additive adversarial perturbations. Future work may consider the effects of adversarial transferability in more invasive AMC environments, where the adversary's knowledge level may be unknown or unpredictable. 

\bibliography{references}

\bibliographystyle{IEEEtran}

\end{document}